\documentclass{svproc}
\usepackage{graphicx}        
\usepackage{amsmath}
\usepackage{color}
\usepackage[normalem]{ulem}
\usepackage{url}

\begin{document}
\mainmatter              
\title{Probing NSI in Atmospheric Neutrino Experiments using Oscillation Dip and Valley}

\titlerunning{Probing NSI using Oscillation Dip and Valley}  
%
\author{Anil Kumar\footnote{Anil Kumar (anil.k@iopb.res.in, ORCID: 0000-0002-8367-8401) would like to thank the organizers of XXIV DAE-BRNS High Energy Physics (HEP) Symposium 2020 for providing an opportunity to present this work.}\inst{1,2,3} \and Amina Khatun\inst{4} \and Sanjib Kumar Agarwalla\inst{1,3,5} \and Amol Dighe\inst{6}}

\authorrunning{Anil Kumar et al.} 
%
\tocauthor{Anil Kumar, Amina Khatun, Sanjib Kumar Agarwalla, and Amol Dighe}

\institute{\textsuperscript{1} IOP, Bhubaneswar, India, \textsuperscript{2} SINP, Kolkata, India, \textsuperscript{3} HBNI, Mumbai, India, \\ \textsuperscript{4} Comenius University, Bratislava, Slovakia, \textsuperscript{5} ICTP, Trieste, Italy, and \textsuperscript{6} TIFR, Mumbai, India}

\maketitle              

\begin{abstract}
We propose a new approach to probe neutral-current non-standard neutrino interaction parameter $\varepsilon_{\mu\tau}$ using the oscillation dip and oscillation valley. Using the simulated ratio of upward-going and downward-going reconstructed muon events at the upcoming ICAL detector, we demonstrate that the presence of non-zero $\varepsilon_{\mu\tau}$ would result in the shift in the dip location as well as the bending of the oscillation valley. Thanks to the charge identification capability of ICAL, the opposite shifts in the locations of oscillation dips as well as the contrast in the curvatures of oscillation valleys for $\mu^-$ and $\mu^+$ is used to constrain $|\varepsilon_{\mu\tau}|$ at 90\% C.L. to about 2\% using 500 kt$\cdot$yr exposure. Our procedure incorporates statistical fluctuations, uncertainties in oscillation parameters, and systematic errors. 

\end{abstract}

In this talk, we propose a new approach to probe neutral-current Non-Standard Interactions (NSI)~\cite{Wolfenstein:1977ue,Farzan:2017xzy} parameter $\varepsilon_{\mu\tau}$ during propagation of atmospheric neutrinos. Due to the charge identification capability, the 50 kt Iron Calorimeter (ICAL) detector at the proposed India-based Neutrino observatory \cite{Kumar:2017sdq} would be able to detect atmospheric neutrinos and antineutrinos separately in the multi-GeV range of energy over a wide range of baselines. The oscillation dip and valley features in the $\nu_\mu$ survival probability can be reconstructed separately for $\mu^-$ and $\mu^+$ using the ratio of upward-going (U) and downward-going (D) reconstructed muon events at the ICAL detector as demonstrated in Ref.~\cite{Kumar:2020wgz}. 

Using the $\chi^2$ analysis with reconstructed muon momentum, it has been demonstrated that one can obtain a bound of $|\varepsilon_{\mu\tau}| < 0.015$ at 90\% C.L.~\cite{Choubey:2015xha} with 500 kt$\cdot$yr exposure at ICAL.  The approach presented in this work is complementary, which uses the observation that, in the presence of non-zero $\varepsilon_{\mu\tau}$, the oscillation dips get shifted in opposite directions for $\mu^-$ and $\mu^+$. In addition, we demonstrate how the contrast in the curvatures of oscillation valleys for $\mu^-$ and $\mu^+$ can also be used to constrain $\varepsilon_{\mu\tau}$.

\section{Shift of the oscillation dip location}
\begin{figure}
	\centering
	\includegraphics[width=0.45\linewidth]{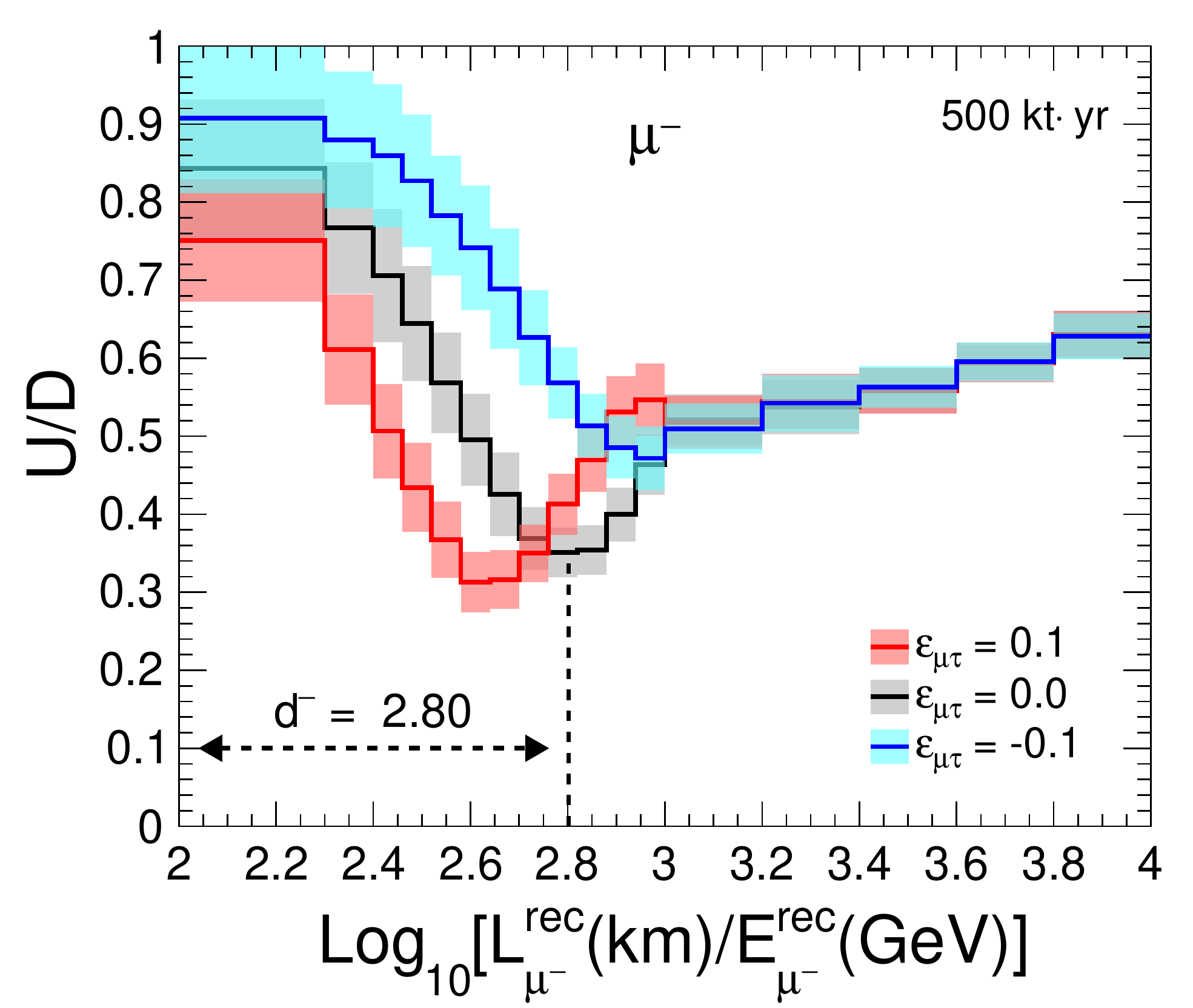}
	\includegraphics[width=0.45\linewidth]{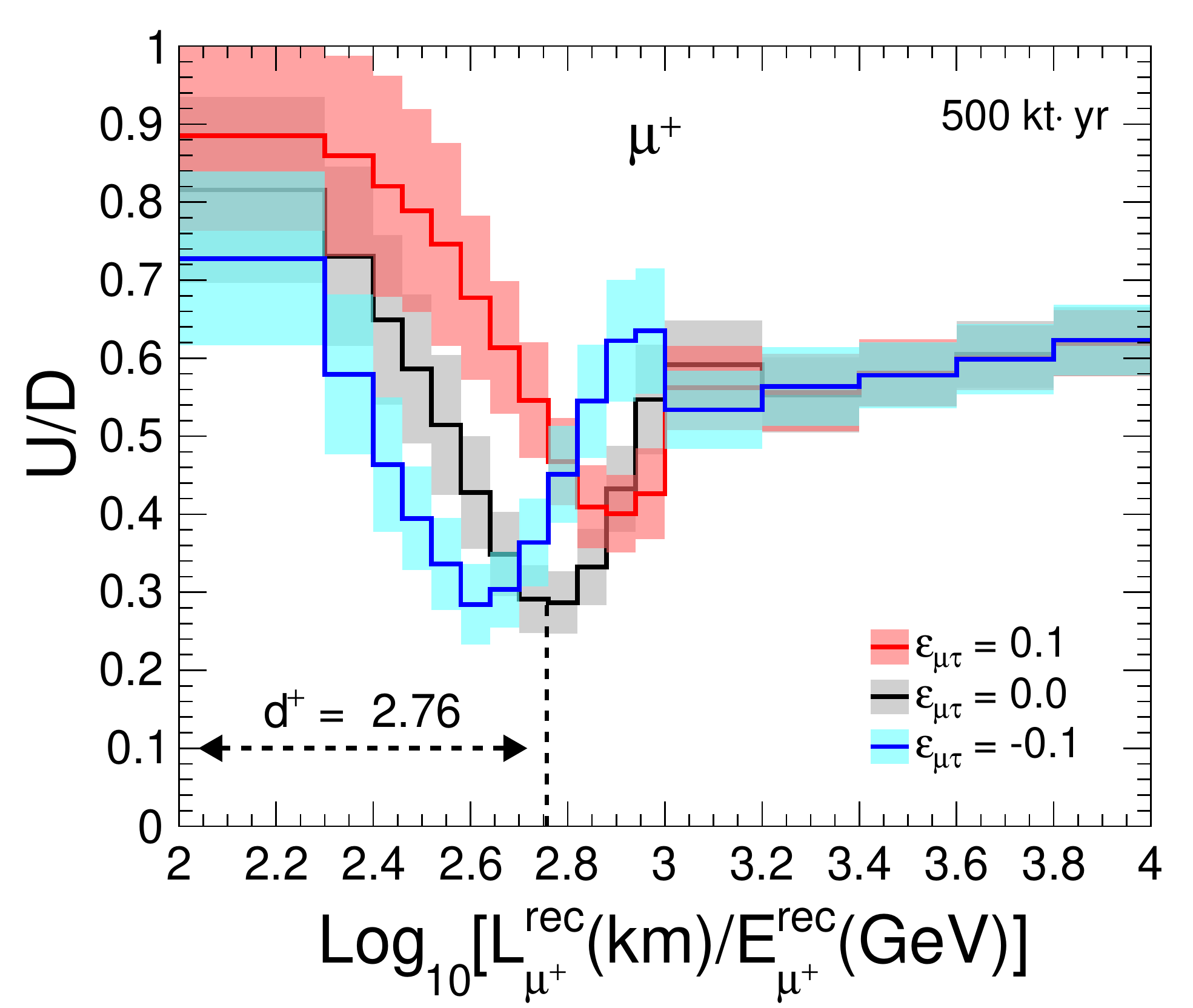}
	\caption{The U/D distributions as a function of $\log_{10}(L_\mu^\text{rec}/E_\mu^\text{rec})$ using simulated sets of 10-year data. The red, black, and blue curves correspond to $\varepsilon_{\mu\tau}$ of 0.1, 0.0, and -0.1, respectively, whereas the colored boxes show statistical fluctuations calculated using 100 simulated sets. Left and right panels correspond to $\mu^-$ and $\mu^+$, respectively. These figures are taken from Ref.~\cite{Kumar:2021lrn}}
	\label{fig:osc_dip}
\end{figure}

We simulate reconstructed muon events at the ICAL detector using NUANCE neutrino event generator, three flavor neutrino oscillations with matter effect considering PREM Profile, and detector properties~\cite{Kumar:2020wgz}. Figure~\ref{fig:osc_dip} shows the oscillation dip in the U/D distribution as a function of $\log_{10}(L_\mu^\text{rec}/E_\mu^\text{rec})$  for a simulated set of 10-year data using $\sin^2 2\theta_{12} = 0.855$, $\sin^2 \theta_{23} = 0.5$, $\sin^2 2\theta_{13} = 0.0875$, $\Delta m^2_{32} = 2.46 \times 10^{-3} ~ (\text{eV}^2)$, $\Delta m^2_{21} = 7.4 \times 10^{-5} ~ (\text{eV}^2)$, and $\delta_\text{CP} = 0$ with normal ordering (NO, $m_1 < m_2 < m_3$). The solid lines show the mean of 100 simulated sets of 10-year data, and the colored boxes show the statistical fluctuations. The red, black, and blue curves are for $\varepsilon_{\mu\tau}$ of 0.1, 0.0, and -0.1, respectively. We can observe that the red (blue) curve shifts towards the left (right) for $\mu^-$ and towards the right (left) for $\mu^+$. We propose a new observable $\Delta d = d^- - d^+$ where $d^-$ and $d^+$ represent the dip locations obtained using dip identification algorithm \cite{Kumar:2020wgz} for $\mu^-$ and $\mu^+$, respectively. The observable $\Delta d$ depends on the value of $\varepsilon_{\mu\tau}$ but is independent of $\Delta m^2_{32}$~\cite{Kumar:2021lrn}. 

\section{Bending of the oscillation valley}
In the plane of energy and direction of neutrino, the oscillation dip feature appears as a straight diagonal band in the absence of NSI, and this band is defined as ``oscillation valley''~\cite{Kumar:2020wgz}. In the presence of non-zero $\varepsilon_{\mu\tau}$, the oscillation valley bends. For the same value of $\varepsilon_{\mu\tau}$, the bending is in opposite directions for neutrino and anti-neutrino. The direction of bending depends on the sign of $\varepsilon_{\mu\tau}$~\cite{Kumar:2021lrn}. 

\begin{figure}
	\centering
	\includegraphics[width=0.45\linewidth]{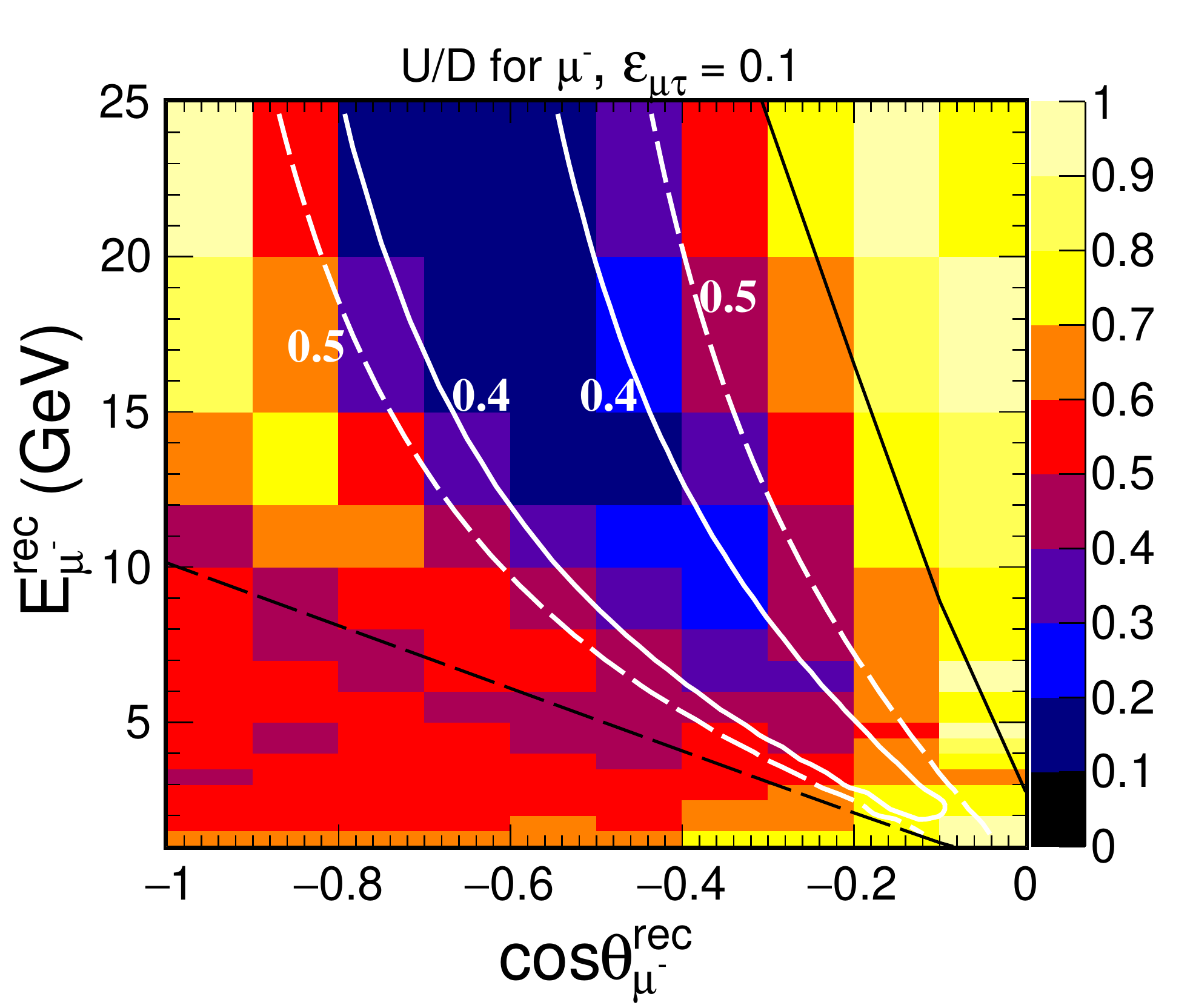}
	\includegraphics[width=0.45\linewidth]{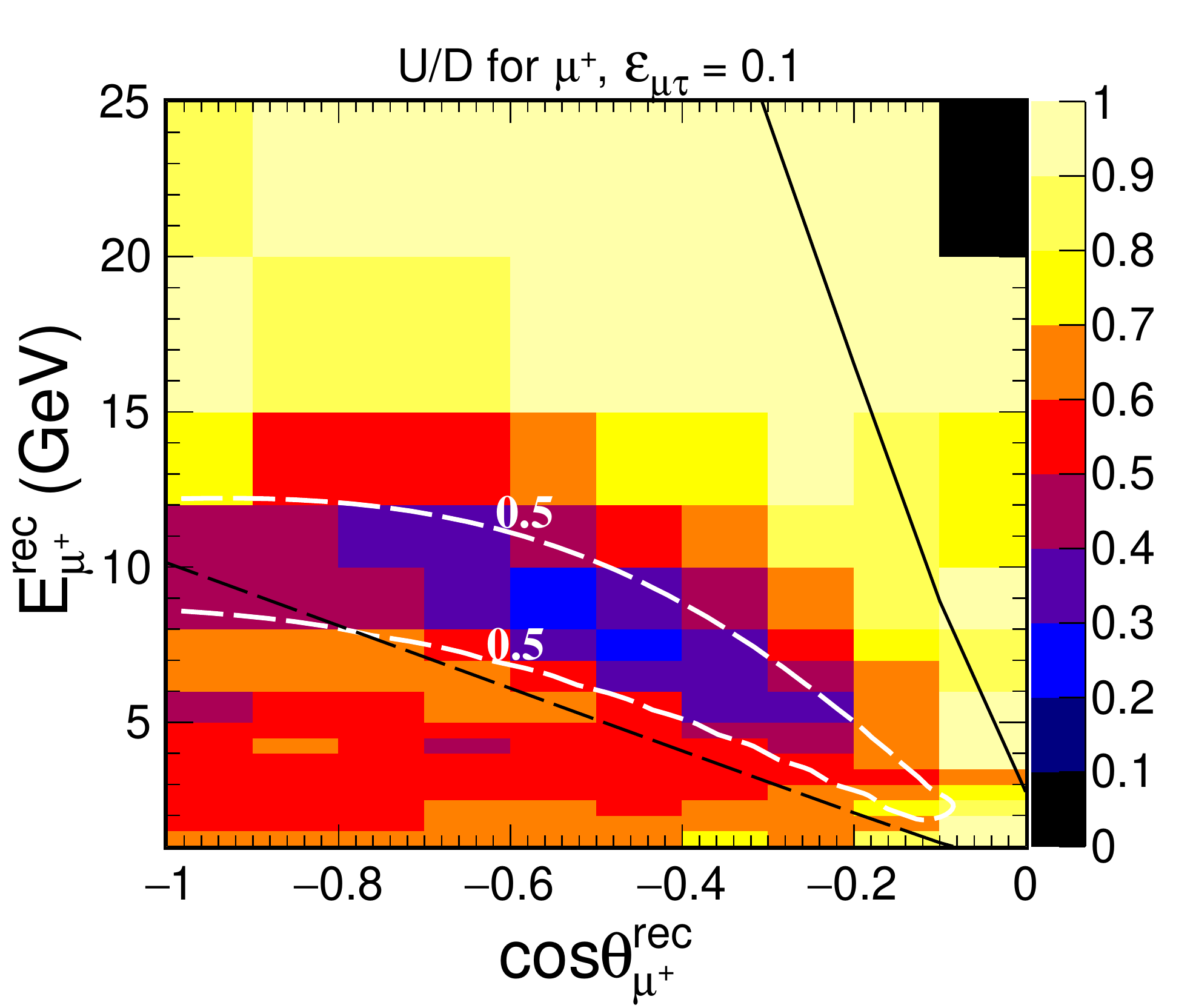}
	\caption{The mean U/D distribution of 100 simulated sets of 10-year data in the plane of ($E_\mu^\text{rec},\cos\theta_\mu^\text{rec}$) of reconstructed muons with $\varepsilon_{\mu\tau} = 0.1$. The white solid and dashed curves represent the fitted function $F_\alpha(E_\mu^{\rm rec}, \cos\theta_\mu^{\rm rec})$ with values 0.4 and 0.5, respectively. Left and right panels correspond to $\mu^-$ and $\mu^+$ respectively. These figures are taken from Ref.~\cite{Kumar:2021lrn}}
	\label{fig:oscillation-valley}
\end{figure}

Figure~\ref{fig:oscillation-valley} shows the mean U/D distributions of 100 simulated sets of 10-year data in the plane of ($E_\mu^\text{rec},\cos\theta_\mu^\text{rec}$) with $\varepsilon_{\mu\tau} = 0.1$ for $\mu^-$ and $\mu^+$. The dark blue diagonal band corresponds to the oscillation valley, which is observed to bend in opposite directions for $\mu^-$ and $\mu^+$. The direction of bending will be reversed for $\varepsilon_{\mu\tau} = -0.1$~\cite{Kumar:2021lrn}. We fit the oscillation valley with the function
\begin{equation}
F_\alpha(E_\mu^{\rm rec}, \cos\theta_\mu^{\rm rec}) = Z_\alpha + N_\alpha \cos^2
\left(m_\alpha \frac{\cos\theta_\mu^{\rm rec}}{E_\mu^{\rm rec}}  + \alpha\,
\cos^2\theta_\mu^{\rm rec}  \right),
\label{eq:fitting-func-nsi} 
\end{equation}
where $Z_\alpha$, $N_\alpha$, $m_\alpha$, and $\alpha$ are free parameters which will be determined from the fitting of the U/D ratio in the plane of ($E_\mu^\text{rec},\cos\theta_\mu^\text{rec}$) as described in Ref.~\cite{Kumar:2021lrn}. The parameters $m_\alpha$ and $\alpha$ contain information about the alignment and the curvature of oscillation valley. The white lines in Fig.~\ref{fig:oscillation-valley} show the contours for representative values of the function $F_\alpha(E_\mu^{\rm rec}, \cos\theta_\mu^{\rm rec})$, which clearly identify 
the curvature of the oscillation valley.

\section{Results}

\begin{figure}
	\centering
	\includegraphics[width=0.45\linewidth]{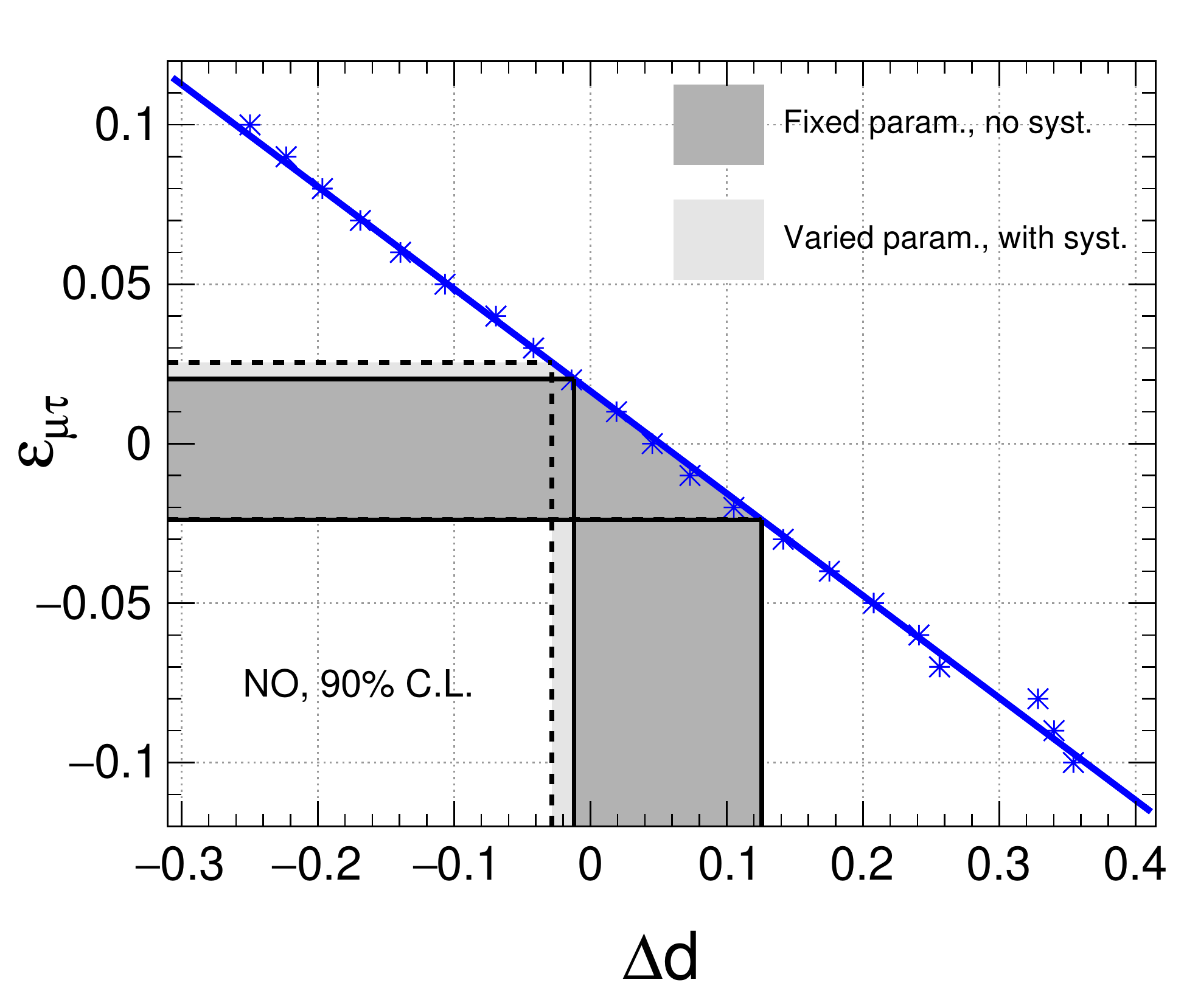}
	\includegraphics[width=0.45\linewidth]{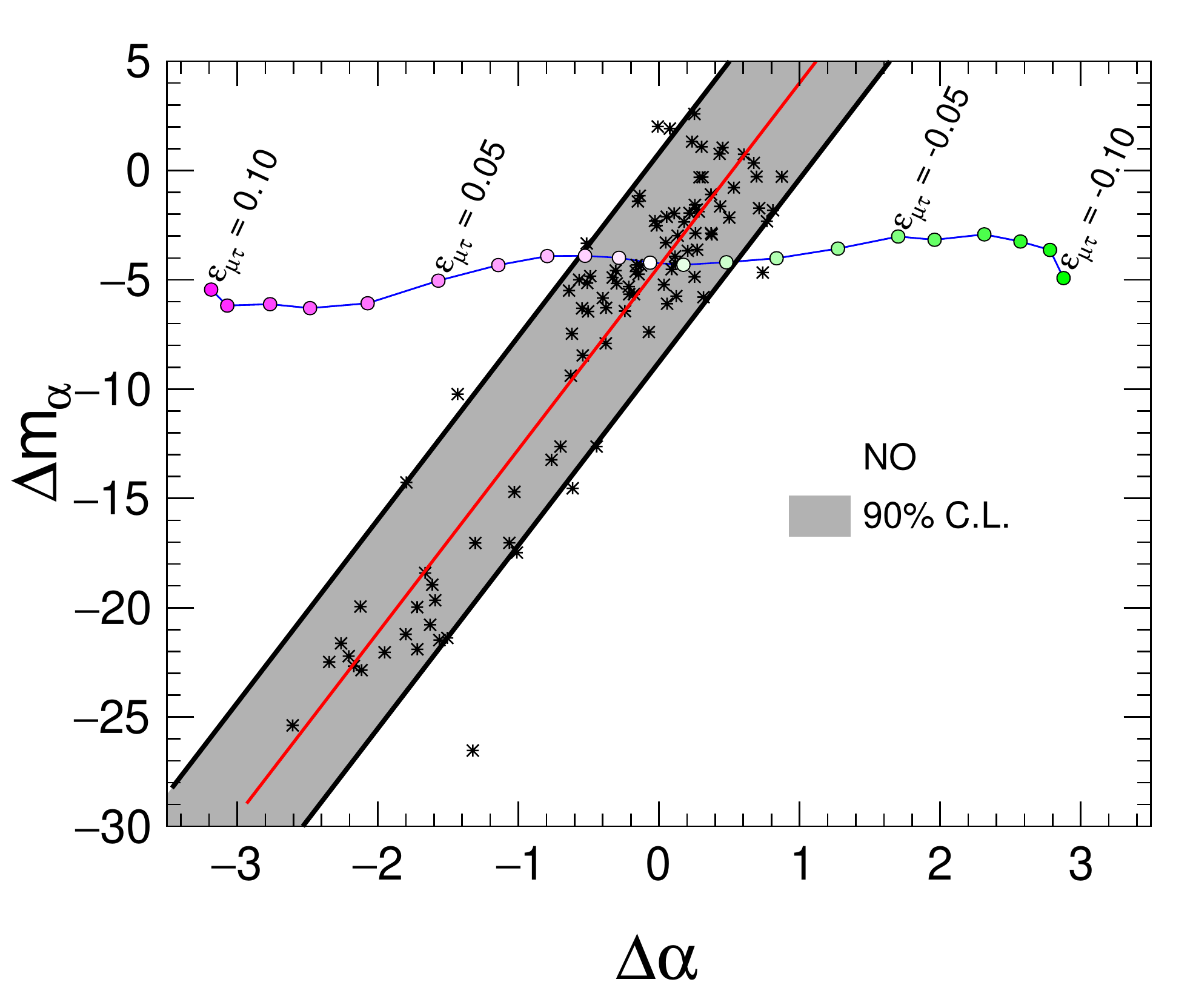}
	\caption{The 90\% C.L. bounds on $\varepsilon_{\mu\tau}$ using oscillation dip (left panel) and oscillation valley (right panel). The blue lines show the calibration curve using 1000-year MC, whereas the dark (light) gray band shows the 90\% C.L. interval obtained using multiple 10-year simulated data sets without (with) variation over oscillation parameters and systematic errors. These figures are taken from Ref.~\cite{Kumar:2021lrn}}
	\label{fig:results}
\end{figure}

 We calibrate $\varepsilon_{\mu\tau}$ with respect to $\Delta d$ using 1000-year Monte Carlo (MC) as shown in the left panel of Fig.~\ref{fig:results} by blue points. We use 100 statistically independent simulated sets of 10-year data with $\varepsilon_{\mu\tau} = 0$, to determine expected bounds on $\Delta d$, and hence on $\varepsilon_{\mu\tau}$. The results with fixed oscillation parameters and without systematic errors, as shown by the dark gray band, give $-0.024 < \varepsilon_{\mu\tau} < 0.020$ at 90\% C.L.. We further incorporate uncertainties in the neutrino oscillation parameters and systematic errors following the procedure mentioned in Ref.~\cite{Kumar:2021lrn} to obtain  the 90\% C.L. bounds on $\varepsilon_{\mu\tau}$ to be $-0.025 < \varepsilon_{\mu\tau} < 0.024$, as shown by the light gray band in Fig.~\ref{fig:results}.

The blue line with colored circles in the right panel of Fig.~\ref{fig:results} shows the calibration curve using 1000-year MC for $\varepsilon_{\mu\tau}$ in the plane of ($\Delta m_\alpha, \Delta\alpha$) where $\Delta \alpha = \alpha^- - \alpha^+$ and $\Delta m_\alpha = m_{\alpha^-} - m_{\alpha^+}$. The black points are obtained after fitting multiple 10-year simulated data sets with $\varepsilon_{\mu\tau} = 0$. The calibration curve overlapped by the gray band gives the expected 90\% C.L. bound for $\varepsilon_{\mu\tau}$, which is $-0.022 < \varepsilon_{\mu\tau} < 0.021$. Variation in oscillation parameters and systematic uncertainties do not affect these results appreciably.

\section{Conclusion}

We demonstrated that the presence of non-zero NSI parameter $\varepsilon_{\mu\tau}$ results in the shift of the oscillation dip location, and the curvature of the oscillation valley. For a given non-zero value of $\varepsilon_{\mu\tau}$, the oscillation dips have opposite shifts, and the oscillation valleys have opposite curvatures for the reconstructed $\mu^-$ and $\mu^+$ events. Thanks to the charge identification capability of ICAL, these features 
can be used to constrain $|\varepsilon_{\mu\tau}|$ at 90\% C.L. to about 2\% using 500 kt$\cdot$yr exposure at ICAL.

\textbf{Acknowledgements:} We acknowledge financial support from the DAE, DST (Govt. of India), and INSA.



\end{document}